\documentclass[aps, prb, reprint, superscriptaddress, 
]{revtex4-2}

\usepackage{graphicx}
\usepackage{dcolumn}
\usepackage{bm}

\usepackage{amsmath, amssymb, graphicx, hyperref, csquotes}

\newcommand{\gk}{$\bar{\mathit{\varGamma}}\!\bar{K}$}
\newcommand{\gm}{$\bar{\mathit{\varGamma}}\!\bar{M}$}

\newcommand{\kgm}{$\bar{K}\!\bar{\mathit{\varGamma}}\!\bar{M}$}
\newcommand{\mgm}{$\bar{M'}\!\bar{\mathit{\varGamma}}\!\bar{M}$}

\newcommand{\mbt}{$\mathrm{MnBi}_2 \mathrm{Te}_4$}

\newcommand{\sbt}{$\mathrm{SnBi}_2 \mathrm{Te}_4$}

\newcommand{\msbt}{$\mathrm{Mn}_{1-x} \mathrm{Sn}_x \mathrm{Bi}_2 \mathrm{Te}_4$}
\newcommand{\mgbt}{$\mathrm{Mn}_{1-x} \mathrm{Ge}_x \mathrm{Bi}_2 \mathrm{Te}_4$}

\begin{document}

\preprint{APS/123-QED}

\title{\textbf{Probing the Interaction Between Topological and Rashba-like Surface States in {\mbt} Through Sn Doping} 
}

\author{A.V.~Tarasov}
\email{artem.tarasov@spbu.ru}
\affiliation{Saint Petersburg State University, 198504 Saint Petersburg, Russia}
\affiliation{Center for Advanced Mesoscience and Nanotechnology, Moscow Institute of Physics and Technology, 141700 Dolgoprudny, Russia}

\author{D.A.~Estyunin}
\affiliation{Saint Petersburg State University, 198504 Saint Petersburg, Russia}
\affiliation{Center for Advanced Mesoscience and Nanotechnology, Moscow Institute of Physics and Technology, 141700 Dolgoprudny, Russia}

\author{A.G.~Rybkin}
 \affiliation{Saint Petersburg State University, 198504 Saint Petersburg, Russia}

\author{A.S.~Frolov}
\affiliation{Center for Advanced Mesoscience and Nanotechnology, Moscow Institute of Physics and Technology, 141700 Dolgoprudny, Russia}
\affiliation{Lomonosov Moscow State University, 119991 Moscow, Russia}

\author{A.I.~Sergeev}
\affiliation{Center for Advanced Mesoscience and Nanotechnology, Moscow Institute of Physics and Technology, 141700 Dolgoprudny, Russia}
\affiliation{Lomonosov Moscow State University, 119991 Moscow, Russia}

\author{A.V.~Eryzhenkov}
\affiliation{Saint Petersburg State University, 198504 Saint Petersburg, Russia}

\author{V.V.~Anferova}
\affiliation{Saint Petersburg State University, 198504 Saint Petersburg, Russia}

\author{T.P.~Estyunina}
\affiliation{Saint Petersburg State University, 198504 Saint Petersburg, Russia}

\author{D.A.~Glazkova}
\affiliation{Saint Petersburg State University, 198504 Saint Petersburg, Russia}

\author{K.A.~Kokh}
    \affiliation{Saint Petersburg State University, 198504 Saint Petersburg, Russia}
    \affiliation{Sobolev Institute of Geology and Mineralogy, Siberian Branch, Russian Academy of Sciences, 630090 Novosibirsk, Russia}

\author{V.A.~Golyashov}
    \affiliation{Saint Petersburg State University, 198504 Saint Petersburg, Russia}
    \affiliation{Rzhanov Institute of Semiconductor Physics, Siberian Branch, Russian Academy of Sciences, 630090 Novosibirsk, Russia}
    \affiliation{Department of Physics, Novosibirsk State University, 630090 Novosibirsk, Russia}

\author{O.E.~Tereshchenko}
    \affiliation{Saint Petersburg State University, 198504 Saint Petersburg, Russia}
    \affiliation{Rzhanov Institute of Semiconductor Physics, Siberian Branch, Russian Academy of Sciences, 630090 Novosibirsk, Russia}

\author{S.~Ideta}
\affiliation{Research Institute for Synchrotron Radiation Science (HiSOR), Hiroshima University, Hiroshima 739-0046, Japan}

\author{Y.~Miyai}
     \affiliation{Research Institute for Synchrotron Radiation Science (HiSOR), Hiroshima University, Hiroshima 739-0046, Japan}

\author{Y.~Kumar}
     \affiliation{Graduate School of Advanced Science and Engineering, Hiroshima University, Higashi-Hiroshima 739-8526, Japan}

\author{K.~Shimada}
    \affiliation{Research Institute for Synchrotron Radiation Science (HiSOR), Hiroshima University, Hiroshima 739-0046, Japan}
    \affiliation{International Institute for Sustainability with Knotted Chiral Meta Matter $(\text{WPI-SKCM}^2)$, Hiroshima University, 739-8526 Higashi-Hiroshima, Japan}
    \affiliation{Research Institute for Semiconductor Engineering, Hiroshima University, 739-8527 Higashi-Hiroshima, Japan}

\author{A.M.~Shikin}
    \affiliation{Saint Petersburg State University, 198504 Saint Petersburg, Russia}    

\date{\today}

\begin{abstract}
The presence of Rashba-like surface states (RSS) in the electronic structure of topological insulators (TIs) has been a longstanding topic of interest due to their significant impact on electronic and spin structures. In this study, we investigate the interaction between topological and Rashba-like surface states (TSS and RSS) in \msbt{} systems using density functional theory (DFT) calculations and high-resolution ARPES. Our findings reveal that increasing Sn concentration shifts RSS downward in energy, enhancing their influence on the electronic structure near the Fermi level. ARPES validates these predictions, capturing the evolution of RSS and their hybridization with TSS. Orbital analysis shows RSS are localized within the first three Te--Bi--Te trilayers, dominated by $Bi-p$ orbitals, with evidence of the orbital Rashba effect enhancing spin-momentum locking. At higher Sn concentrations, RSS penetrate deeper into the crystal, driven by Sn $p$-orbital contributions. These results position \msbt{} as a tunable platform for tailoring electronic properties in spintronic and quantum technologies.
\end{abstract}

\maketitle


\section{\label{sec:intro}Introduction}

Topological quantum materials stand at the forefront of modern physics, captivating researchers with their remarkable electronic phenomena, among which topological surface states (TSS) play a central role~\cite{hasan_colloquium_2010, yan_topological_2012, zhang_topological_2009}. These states, arising from the unique topology of the electronic band structure, exemplify the novel and often counterintuitive behaviors that define this exciting class of materials. These spin-polarized states emerge at the interface between topological and trivial phases and are protected by time-reversal symmetry (TRS). This protection ensures their robustness against local perturbations and prevents electron backscattering, maintaining their unique conductive properties. Moreover, introducing magnetic interactions into such systems can open a gap in TSS, enabling phenomena such as the quantum anomalous Hall effect (QAHE), the topological magnetoelectric effect (TME), and other unique topological properties~\cite{qi2008topological, qi2011topological, tokura2019magnetic, chang2023colloquium, chang2013experimental, wang2021intrinsic} This development has given rise to a distinct field of research focused on magnetic topological insulators, expanding the horizons of topological quantum materials.

Intrinsic magnetic topological insulators (TI), such as \mbt{}, represent a class of materials where magnetic atoms occupy definite crystallographic positions \cite{otrokov2019prediction, li2019intrinsic, zhang2019topological, gong2019experimental}. This regular arrangement enables quantum topological effects like QAHE to be observed at higher temperatures, with \mbt{} thin films demonstrating QAHE at 1.4K without external magnetic fields or at 6.5K with applied fields \cite{deng2020quantum}.

Shortly after the first publication on the experimental investigation of this material~\cite{otrokov2019prediction}, reports emerged of an additional parabolic conduction band states observed in ARPES data~\cite{estyunin2020signatures}. These states, referred to in that work as Rashba-like surface states (RSS), were noted to exhibit lifting of the degeneracy at the Kramers point and the emergence of a band gap upon crossing the Néel temperature, consistent with the behavior of RSS in systems with magnetism.

Interestingly, the ab initio calculations available at that time~\cite{otrokov2019prediction, hao_gapless_2019, swatek_gapless_2020} did not reproduce these states.
Subsequently, in ref.~\cite{nevola_coexistence_2020}, a circular dichroism study firmly established the band dispersion of these states, providing strong evidence of spin-momentum locking, as expected for RSS. Later, in ref.~\cite{liang_approaching_2022}, it was revealed that a topologically nontrivial TSS and trivial RSS coexist and hybridize near the Fermi energy in pristine \mbt{}. Using high-resolution laser- and synchrotron-based angle-resolved photoemission spectroscopy (ARPES), it was demonstrated that the formation of random K clusters and alloying effects significantly modify the surface potential, ultimately suppressing the RSS and leaving only the TSS and inverted bulk bands. This process brings the system closer to the minimal topological electronic band structure characteristic of a magnetic topological insulator (MTI).

A renewed interest in the coexistence of RSS and TSS in MTI has emerged from studies of systems like $(\mathrm{Mn}_{1-x}A_x^4)\mathrm{Bi}_2\mathrm{Te}_4$, where $A^4 = \mathrm{Ge}, \mathrm{Pb}, \mathrm{Sn}$. It was found that, as the concentration of $A^4$ increases, the contribution of RSS to the electronic structure becomes increasingly pronounced in comparison to pristine MBT~\cite{tarasov2023topological, estyunina2023evolution, frolov2024magnetic, frolov2024magnetic, shikin2024phase, estyunin2024pb}. Recent studies on systems where $A^4$ is $\mathrm{Ge}$ or $\mathrm{Pb}$ have demonstrated that RSS significantly influences the spin texture of these materials~\cite{shikin2024spin, estyunin2024pb}. Unlike the RSS observed in pristine $\mathrm{MnBi}_2\mathrm{Te}_4$, the RSS in $(\mathrm{Mn}_{1-x}A_x^4)\mathrm{Bi}_2\mathrm{Te}_4$ systems are well-reproduced by DFT calculations, as confirmed in recent works~\cite{tarasov2023topological, estyunina2023evolution, frolov2024magnetic, shikin2024phase}. This makes these systems an excellent platform for exploring the interplay between RSS and TSS, offering valuable insights into their mutual influence and potential applications.

One of the promising research targets in this domain is the \msbt{} system, a solid solution of the magnetic topological insulator \mbt{} and the nonmagnetic topological insulator \sbt{}. Earlier investigations of these systems revealed that the complex hybridization of $\mathrm{Te}$-$p_{z}$ and $\mathrm{Bi}$-$p_{z}$ orbitals with the orbitals of Sn leads to the emergence of various topological phases depending on the Sn concentration~\cite{tarasov2023topological}. This indicates the possibility of topological phase transitions in such systems, making \msbt{} an excellent platform for tuning the behavior of RSS, TSS, and their mutual interactions.

Despite numerous studies reporting the appearance of RSS in the electronic structure of \mbt{}-based systems, their precise origin and characteristics remain elusive. In particular, the Rashba-like behavior of these states has yet to be conclusively verified. By examining Sn-doped systems, this study aims to elucidate the properties of RSS through an in-depth analysis of their interaction with the TSS, shedding light on the evolution of electronic and spin textures in these materials. This work addresses key knowledge gaps and contributes to a deeper understanding of RSS in \mbt-based compounds.

\section{\label{sec:methods}Methods}

First-principles calculations in the framework of the density functional theory
(DFT) were partially performed at the Computing Center of SPbU Research park. The
electronic structure calculations with impurities were conducted using
the OpenMX DFT code which implements a linear combination of pseudo-atomic
orbitals (LCPAO) approach \cite{ozaki2003variationally, ozaki2004numerical,
ozaki2005efficient} with full-relativistic norm-conserving pseudopotentials
\cite{troullier1991efficient}. The GGA-PBE exchange-correlation functional
\cite{perdew1996generalized} was utilized and basis sets were specified as
Sn7.0-s3p2d2, Mn6.0-s3p2d1, Te7.0-s3p2d2f1 and Bi8.0-s3p2d2f1 where numbers mean
interaction ranges in \AA. Real-space numerical integration accuracy was
specified by cutoff energy of 300~Ry, total energy convergence criterion was
$1\times 10^{-6}$ Hartree. The Mn~$3d$ states were considered within the
$\text{DFT}+U$ approach \cite{dudarev1998cj} with $U = 5.4$~eV.

Simulation of \msbt{} compounds for $x$ values of 25\%, 50\% and 75\% was
performed using $2 \times 2$ supercells. The k-mesh for Brillouin zone sampling in the bulk band structure calculations was set to a \( 5 \times 5 \times 5 \) grid. Surface band structures were calculated for 6~SL (septuple-layer) slabs separated by 20~\AA{} vacuum gaps in the $z$ direction (the surface normal). The k-meshes for Surface Brillouin zones were specified as follows: $5 \times 5$ mesh for pristine \mbt{} and \sbt{} slabs and $3 \times 3$ mesh for $2 \times 2$ supercells.

The Effective Screening Medium (ESM) method~\cite{Otani_esm} was employed to account for the impact of an additional surface charge using the "Charged slab: vacuum/metal" boundary condition, and the "Isolated slab: vacuum/vacuum" boundary condition was applied for calculations involving a uniform external electric field represented by a sawtooth waveform.

ARPES measurements (Fig.~\ref{fig5} a1-a4) were performed at the Rzhanov Institute of Semiconductor Physics SB RAS (Novosibirsk, Russia) using the SPECS ProvenX-ARPES facility equipped with the SPECS ASTRAIOS 190 analyser. He I$\alpha$ ($h\nu = 21.2$~eV) radiation and liquid nitrogen cooling were used for the ARPES measurements, while Al K$\alpha$ ($h\nu = 1486.7$~eV) radiation was employed for stoichiometry estimation of the samples via X-ray photoelectron spectroscopy (XPS).

Laser-based experiments were performed at the $\mu$-Laser ARPES station with Scienta R4000 analyzer (Fig.~\ref{fig5} b1–b4) 
at HiSOR (Hiroshima, Japan). A laser beam with photon energy of $h\nu = 6.3$~eV and a photon flux of up to 10$^{14}$ photons/s was used for the $\mu$-ARPES measurements. The expected energy resolution was approximately 5 meV. The incident photon beam spot diameter was estimated to be
5--10~$\mu$m.

During the measurement process, the temperature of the samples was maintained below 30~K. The base pressure in the
analytic vacuum chamber was less than $10^{-10}$~Torr for all performed measurements.

The analysis of diffraction patterns and refinement of unit cell parameters were performed using the Jana2006 software. Polycrystalline samples were examined using a Rigaku SmartLab SE X-ray powder diffractometer in Bragg-Brentano reflection geometry. The measurements employed copper radiation (Cu K$\alpha_1$ + K$\alpha_2$) over a $2\theta$ range of $5^\circ$ to $80^\circ$, with a step size of $0.01^\circ$ and a scan speed of $5^\circ$/min. The primary X-ray beam was monochromatized using a germanium monochromator (111 plane).

Single-crystal samples were grown using the Bridgman method and measured on a Bruker D8 Discover diffractometer. These measurements used copper radiation (Cu K$\alpha_1$) with a germanium monochromator (022 plane) over a $2\theta$ range of $5^\circ$ to $90^\circ$, with a step size of $0.025^\circ$. Polycrystals were synthesized using a solid-state method followed by prolonged annealing.

\section{\label{sec:results}Results and discussion}
\subsection{\label{sec:crystal}Crystal growth and characterization}

\begin{figure*}[ht!]
    \centering
    \includegraphics[width=1.0\textwidth]{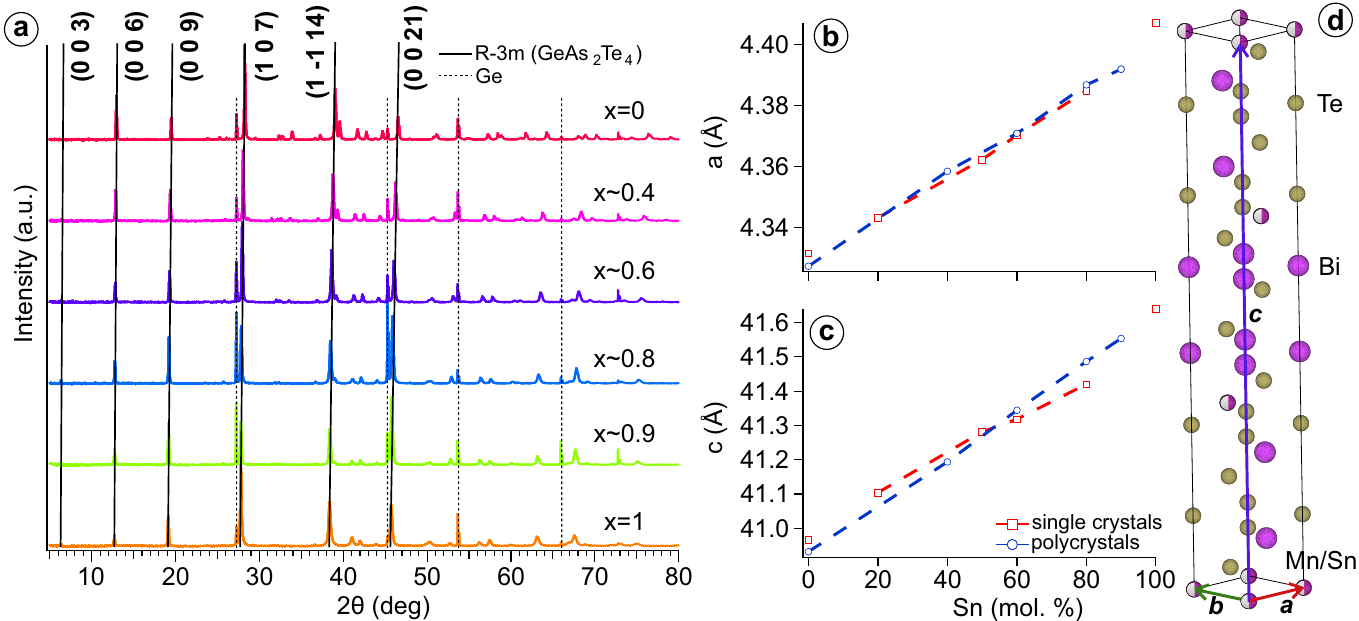}
    \caption{\textbf{(a)} Powder diffraction patterns obtained from polycrystalline samples. The diffraction peaks are labeled, and their positions correspond to the $\mathrm{Ge} \mathrm{As}_2 \mathrm{Te}_4$ structural type.
\textbf{(b, c)} Lattice constants $a$ and $c$ as functions of Sn content ($x$), with blue lines and markers representing data for polycrystalline samples and red lines and markers for single-crystal samples, illustrating a close match between the two types of data. The errors in the refined parameters are proportional to the size of the markers on the plot.
\textbf{(d)} Schematic representation of the \msbt{} crystal structure, highlighting the atomic arrangement and structural symmetry.}
    \label{fig1}
\end{figure*}

Initially, \msbt{} polycrystalline and single crystals samples were grown from the melt using the Bridgman method. Growth was performed from the melt of \msbt{} with x taken equal to the targeted composition. X-Ray Diffraction (XRD) data are collected in Fig.~\ref{fig1}.

The Fig.~\ref{fig1}(a) shows the powder diffraction patterns obtained from polycrystalline samples synthesized by melting elemental components (Fig.~{1S} in Suppl. Inform. represents similiar data for single crystals samples). To determine and refine the unit cell parameters, monocrystalline germanium powder was added in an amount of 10 wt\% relative to the total powder mass. The reflections from germanium are indicated with thin dashed lines, while the most characteristic and intense reflections of the studied samples are labeled and highlighted with bold lines. From the set of interplanar spacings, it can be concluded that all compounds in the \msbt{} system crystallize in the $\mathrm{Ge} \mathrm{As}_2 \mathrm{Te}_4$ structural type. Additionally, a gradual shift in the reflection positions is observed with increasing Sn content, suggesting the progressive substitution of manganese by Sn in the cation positions of the 124 structure while preserving the overall crystal symmetry.

Fig.~\ref{fig1}(b, c) shows the dependencies of the refined lattice parameters on the Sn content (blue lines and markers for polycrystalline samples). These dependencies exhibit a linear trend, indicating adherence to ’s law. Notably, Fig.~\ref{fig1}(b, c) also includes data obtained for single-crystal samples (red lines and markers), which are in close agreement with the results for polycrystalline samples. This near-coincidence further supports the thermodynamic equilibrium and homogeneity of the synthesized materials. Minor deviations from linearity may reflect slight compositional inhomogeneity or deviations from stoichiometry.

\subsection{Electronic band structure calculations}

\begin{figure*}[ht!]
    \centering
    \includegraphics[width=1.0\textwidth]{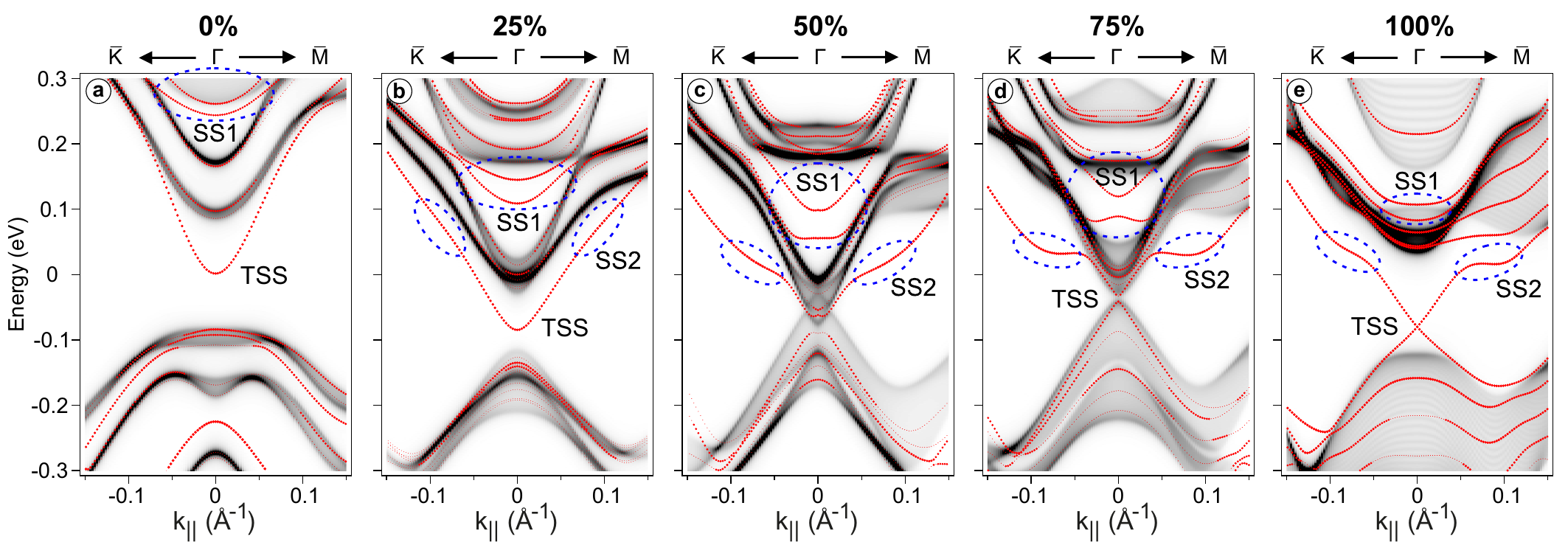}
    \caption{Electronic structure calculations for the nearest valence and conduction band states across different Sn concentrations: \textbf{(a)} 0\%, \textbf{(b)} 25\%, \textbf{(c)} 50\%, \textbf{(d)} 75\%, and \textbf{(e)} 100\%. The TSS, along with the surface states SS1 and SS2 (where observable), are explicitly identified. Bulk-projected bands are depicted as a gray gradient, while red dots denote slab bands with significant surface contributions (exceeding 40\%).}
    \label{fig2}
\end{figure*}

Fig.~\ref{fig2} presents the results of DFT calculations of the electronic structure for \msbt{} systems at various Sn concentrations. The lattice constants of the unit cell for different $x$ values were set according to the linear functions on $x$, determined by analyzing XRD data (Fig.~\ref{fig1}(b, c)). At the same time, the atomic positions within the unit cell are determined as the average coordinates of their locations in pristine crystals of \mbt{}~\cite{mbt_str} and \sbt{}~\cite{sbt_str} according to the parameter $x$, accounting for changes in the crystal structure induced by doping.

In Fig.~\ref{fig2}(a–e), the electronic structure of the slab is overlapped with the bulk states projections integrated along the $k_z$ axis, providing a detailed view of the evolution of surface and bulk states as the Sn concentration varies. Fig.~\ref{fig2}(a) illustrates the electronic structure of \mbt{}, consistent with previously published theoretical calculations for this material~\cite{zhang2019topological,otrokov2019prediction}. The figure highlights topological surface states (TSS) within the bulk band gap, which exhibit a gap associated with the presence of magnetic Mn atoms in the system. Additionally, parabolic surface states (SS1) are visible above the Fermi level within the local bandgap of the bulk band projections.

As the Sn concentration increases to 25\% (Fig.~\ref{fig2}(b)), significant changes are observed in the surface electronic structure. The SS1 shift downward in energy, and a distinct bend emerges in the upper part of the Dirac cone. This deviation, labeled SS2, is characterized by noticeable nonlinearity in the dispersion branches along the \gk{} and \gm{} directions.

Increasing the Sn concentration to 50\% leads to dramatic changes in the electronic structure (Fig.\ref{fig2}(c)). The bulk bandgap completely disappears, eliminating the possibility of TSS existence. This transition, as discussed in previous studies on \msbt{} systems\cite{tarasov2023topological}, is linked to topological phase transitions, suggesting the emergence of an intermediate gapless phase between the two distinct topological phases. A similar gapless state was recently observed experimentally in the \mgbt{} system at $x=0.42$~\cite{frolov2024magnetic}, and similar changes in the electronic structure are observed in ref.~\cite{shikin2024phase}. Notably, for the 50\% Sn system, the disappearance of TSS is accompanied by a pronounced presence of the SS2 feature, indicating that SS2 is not simply a modified form of TSS branches but rather an independent surface states. The energy position of SS2 appears to be influenced by Sn doping, which causes this feature to shift downward in energy.

With a further increase in Sn concentration to 75\% (Fig.~\ref{fig2}(d)), a bandgap reopens in the bulk electronic structure and this allows the reformation of TSS. Additionally, the SS1 become increasingly similar to Rashba states of magnetically doped BiTeI with the energy gap opening at the Kramers point~\cite{Shikin_Bite_2017}, particularly due to the reducing splitting of SS1 at the $\Gamma$ point. Similarly, the dispersion of SS2 also acquires a Rashba-like character, indicating a progressive transformation of its electronic structure towards a spin-split parabolic dispersion.

The evolution of the electronic structure reaches its final stage in pristine \sbt (Fig.~\ref{fig2}(e)). The electronic structure of this material closely resembles that of $\mathrm{Mn}_{0.25} \mathrm{Sn}_{0.75} \mathrm{Bi}_2 \mathrm{Te}_4$, but with notable differences. The TSS, along with SS1 and SS2, remain present but undergo significant modifications. In the absence of magnetism, the gap in the TSS is closed.
The SS1 state regains a parabolic shape, resembling the electronic states observed in a pristine \mbt{} crystal, indicating a return to its intrinsic electronic structure.
Simultaneously, the SS2 state becomes more asymmetric along the \kgm{} path, indicating its continued evolution in response to changes in the electronic structure. 

Thus, Fig.~\ref{fig2} illustrates the gradual evolution of the electronic structure from an antiferromagnetic topological insulator to a nonmagnetic topological insulator as the Sn concentration increases, passing through intermediate phases. This is primarily expressed in the modification of the SS1 and SS2 features, as well as the presence or absence of TSS and the gap within them. These observations highlight the direct influence of Sn doping on the electronic structure and its ability to modify surface state behavior in a controlled manner.

\begin{figure*}[ht!]
    \centering
    \includegraphics[width=1.0\textwidth]{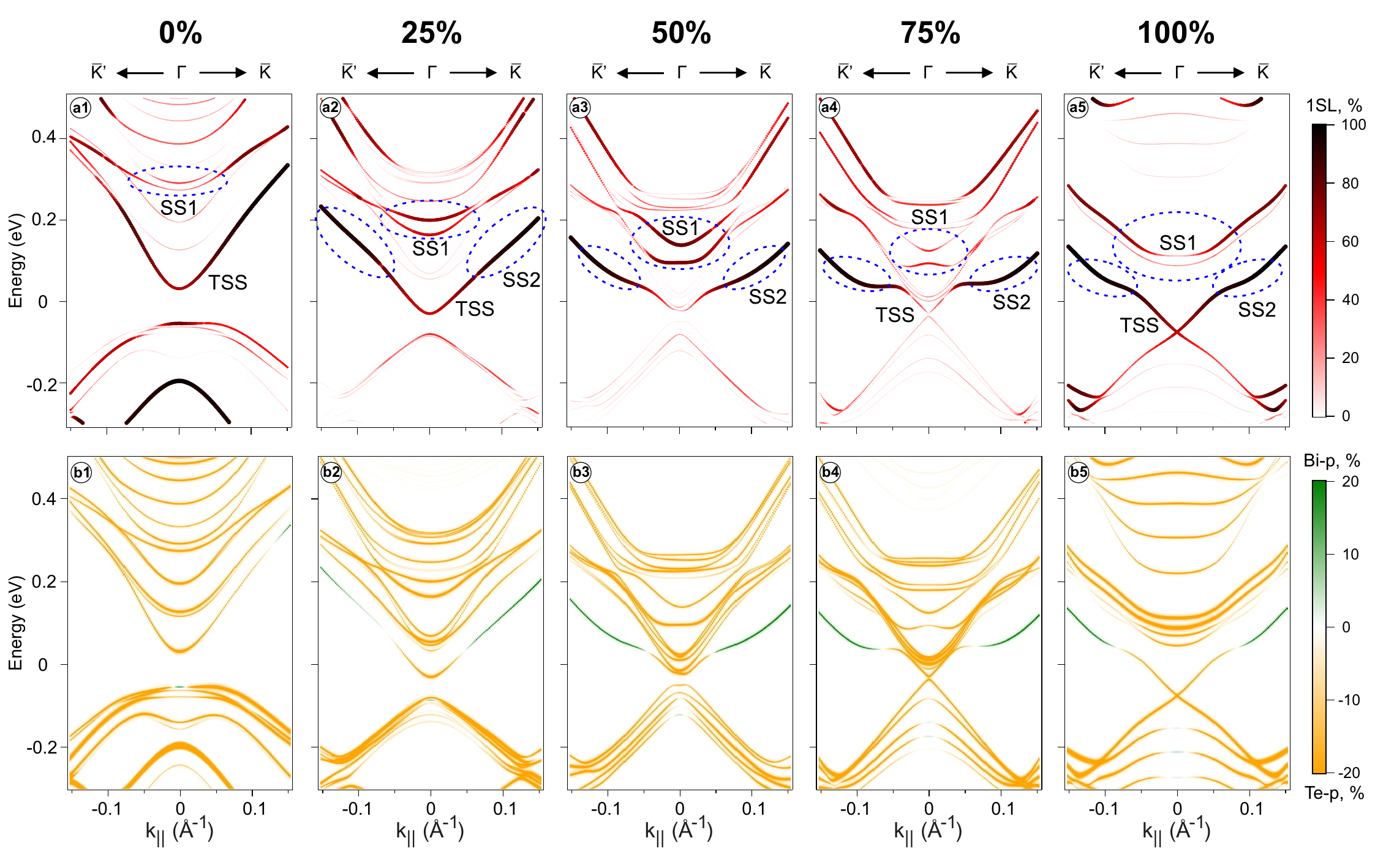}
    \caption{The orbital composition of the nearest valence and conduction band states is shown for varying Sn concentrations: \textbf{(a1, b1)} 0\%, \textbf{(a2, b2)} 25\%, \textbf{(a3, b3)} 50\%, \textbf{(a4, b4)} 75\%, and \textbf{(a5, b5)} 100\%. In the top row \textbf{(a)}, the total contribution of atoms from the first SL block is displayed. The relative surface contribution is highlighted in black-red-white. In the bottom row \textbf{(b)}, the difference between the contributions of $Bi-p$ and $Te-p$ orbitals for atoms in the first SL is shown using a green-white-orange color scale. Green shading indicates the dominance of $Bi-p$ contributions, while orange shading highlights the dominance of $Te-p$ contributions.}
    \label{fig3}
\end{figure*}

The spatial distribution of electronic states within the crystal can also be examined for a more detailed analyzing the interaction between TSS, SS1, SS2 and other overlapping surface states. Fig.~\ref{fig3}(a) illustrates the contribution of atoms from the first septuple layer (SL) block to the electronic structure (Fig.~2S in Suppl. Inform. shows similar data for the second septuple layer and their combined contribution). Notably, the SS2 are almost entirely localized within the first SL for all Sn concentrations. Of particular interest are the SS1, which exhibit a less pronounced surface character that changes significantly with varying $x$. A potential factor influencing their localization is hybridization with other valence states, which evolves as the Sn concentration changes.

For a more detailed analysis of the surface states, let us examine their orbital composition. In Fig.~\ref{fig3}(b), the green regions correspond to areas dominated by the $Bi-p$ contribution, while the orange regions indicate the dominance of $Te-p$ orbitals (for atoms of the first SL). The analysis of this data reveals that the SS2 exhibit a distinct $Bi-p$ character across all Sn concentrations, up to 100\%. This strongly suggests that SS2 originate from the $Bi-p$ orbitals. In contrast, SS1 consistently retain the dominance of $Te-p$ orbitals across all Sn concentrations.

\begin{figure*}[ht!]
    \centering
    \includegraphics[width=\textwidth]{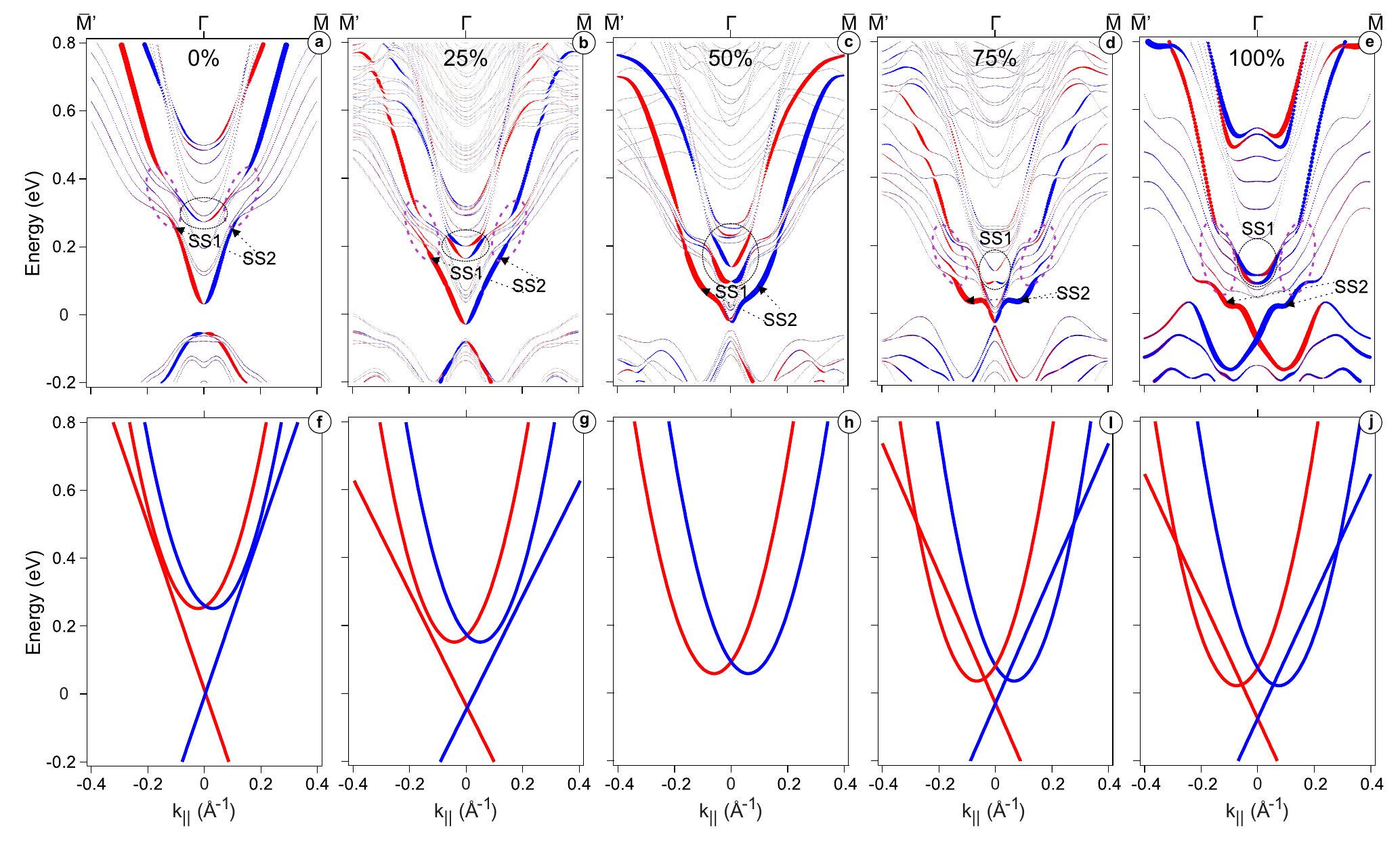}
    \caption{The spin texture ($s_{y}$) of surface states localized in the first SL along the \mgm{} path in the Brillouin zone is shown for various Sn concentrations: (\textbf{a}) 0\%, (\textbf{b}) 25\%, (\textbf{c}) 50\%, (\textbf{d}) 75\%, and (\textbf{e}) 100\%. The \mgm{} direction lies along the \textbf{x}-axis of the Cartesian coordinate system, where the $s_{y}$ spin component is defined (red: positive spin; blue: negative spin). Dashed circles indicate the presumed locations of avoided crossings between the TSS and RSS. The bottom row (\textbf{f-j}) shows the proposed arrangement of TSS and RSS in the absence of hybridization effects.}
    \label{fig4}
\end{figure*}

To understand the origin and characteristics of the SS1 and SS2, we can address to the spin texture of surface states localized in the first SL along the \mgm{} path in the Brillouin zone (Fig.~\ref{fig4}). Fig.{3S} in Suppl. Inform. shows similar data for a narrower range of k-points near the $\Gamma$ point, with a decomposition into contributions from the first and second SLs. It is evident that even at 0\% Sn concentration (Fig.~\ref{fig4}(a)), the spin texture of the studied system cannot be described solely by spin-polarized TSS.
A clear hybridization is observed between TSS and other surface states located above 0.25 eV, which appear to exhibit a parabolic dispersion (becoming more pronounced at higher Sn concentrations) suggesting their Rashba-like origin. We suggest that the branches of the TSS with opposite spin polarization intersect with these surface states at the areas marked with dashed circles in Fig.~\ref{fig4}(a). This is schematically represented in Fig.~\ref{fig4}(f) using branches of TSS and RSS with opposite spin polarization. This visualization highlights the interplay between these distinct types of spin-polarized states, which significantly complicates the spectral structure, causing substantial deviations of the states from their initial dispersions, and leads to noticeable changes in the shape of the TSS and RSS.

Notably, the energy position of the hybridization gaps at 0.3 eV coincides with the position where the SS1 states were previously identified. If we assume that at a distance of 0.2 \AA$^{-1}$ from the $\Gamma$ point, intense "sections" of parabolic states with opposite spin polarization are observed, the SS1 feature can then be described as RSS strongly hybridized with TSS (as schematically shown in Fig.~\ref{fig4}(f)) and surrounding them electronic states. Within the considered energy range, in addition to TSS and SS1, the spin-polarized states emerge at 0.6 eV, which should also be attributed to hybridization effects.

As the Sn concentration increases to 25\% (Fig.~\ref{fig4}(b)), this interpretation becomes even more substantiated. The dispersion of the TSS becomes increasingly difficult to describe with simple Dirac cone (as highlighted in Fig.~\ref{fig4}(g)). Furthermore, the previously observed hybridization gap shifts approximately 0.1 eV lower in energy, while SS1 and other spin-polarized states within the Dirac cone also undergo a downward energy shift. Thus, we can state that, while the main features of the pattern are preserved, the RSS noticeably shift to lower energies, as schematically shown in Fig.~\ref{fig4}(g).

As previously noted, at $x=0.5$, the system transitions into the trivial phase. This transition is clearly reflected in the disappearance of spin-polarized TSS, as observed in Fig.\ref{fig4}(c). Here, the surface spin texture above the Fermi level is formed exclusively by RSS (Fig.\ref{fig4}(h)), with no hybridization with TSS, while the inner branches of the RSS become the most prominent, as well as their hybridization with other surface states, which is particularly evident in the pronounced appearance of SS1. This absence of TSS-RSS hybridization is particularly evident in the disappearance of the hybridization gap, previously observed in Figs.\ref{fig4}(a, b) at energies of 0.3~eV and 0.2~eV, respectively. Notably, the energy positions of the RSS also evolve with increasing Sn concentration, continuing their trend of shifting further downward in energy.

The situation becomes more complex as the system returns to the topological insulator state at $x=0.75$ (Fig.\ref{fig4}(d)). It is worth noting that at 75\% Sn, an indirect gap is observed, similar to the case of 100\% Sn reported in earlier theoretical calculations~\cite{Eremeev2023}, where the maximum of the bulk valence band is located along the \gm{} direction of the BZ. The spectrum clearly reveals the emergence of the upper cone of the TSS. The inner branches of the RSS become barely distinguishable due to their overlap with other electronic states, while the outer branches remain prominently visible. Notably, the hybridization gap reappears at 0.1~eV, further confirming the interaction between TSS and RSS (Fig.\ref{fig4}(i)). The downward shift of the RSS in energy noticeably slows compared to previous transitions.

As the Sn concentration increases to 100\% (Fig.~\ref{fig4}(e, j)), transitioning the system into a nonmagnetic topological insulator state, the overall behavior of the outer branches of the RSS remains largely similar to the case at 75\% Sn. The most notable differences include the hybridization of the lower TSS cone with valence band states and the emergence of intense spin-polarized states at 0.6~eV, which exhibit a clear spin-momentum locking effect. Interestingly that similar states at this energy are also observed for the pristine \mbt{} (Fig.\ref{fig4}(a)). This emergence is likely associated with the complex hybridization between the inner branches of the RSS and the TSS. The Fermi level shifts upward in energy relative to the Dirac point, making it impossible to accurately estimate the energy shift of the RSS compared to other concentrations. Additionally, the hybridization gap within the RSS becomes significantly wider.

The presented analysis highlights the intricate evolution of the electronic structure in \msbt{} as the Sn concentration varies, showcasing the interplay between RSS, TSS, and their hybridization effects. The doping-induced transitions from an antiferromagnetic to a nonmagnetic topological insulator state, passing through intermediate phases, are marked by significant modifications in the SS1 and SS2 features, the reformation or disappearance of TSS, and shifts in hybridization gaps. Notably, the downward energy shifts of RSS with increasing Sn concentration emphasize the role of Sn doping in tailoring the surface state behavior. These observations provide a foundation for further understanding the mechanisms driving surface state evolution and their dependence on doping concentration.

\subsection{Photoemission analysis of electronic structure}
To validate our theoretical results, we utilized helium lamp ($h\nu$ = 21.2 \text{eV}) and laser ($h\nu$ = 6.3\ \text{eV}) angle-resolved photoemission spectroscopy (ARPES) measurements.
The differences in the excitation energy and intensity of the radiation enable selective analysis of states with varying depths of localization and photoemission cross-sections, providing complementary insights into the electronic structure of the system. This is particularly relevant for our study, as the TSS could only be observed at photon energies below 18~eV, likely due to matrix element and cross-section effects~\cite{chen2019topological}.

\begin{figure*}[ht!]
    \centering
    \includegraphics[width=16cm]{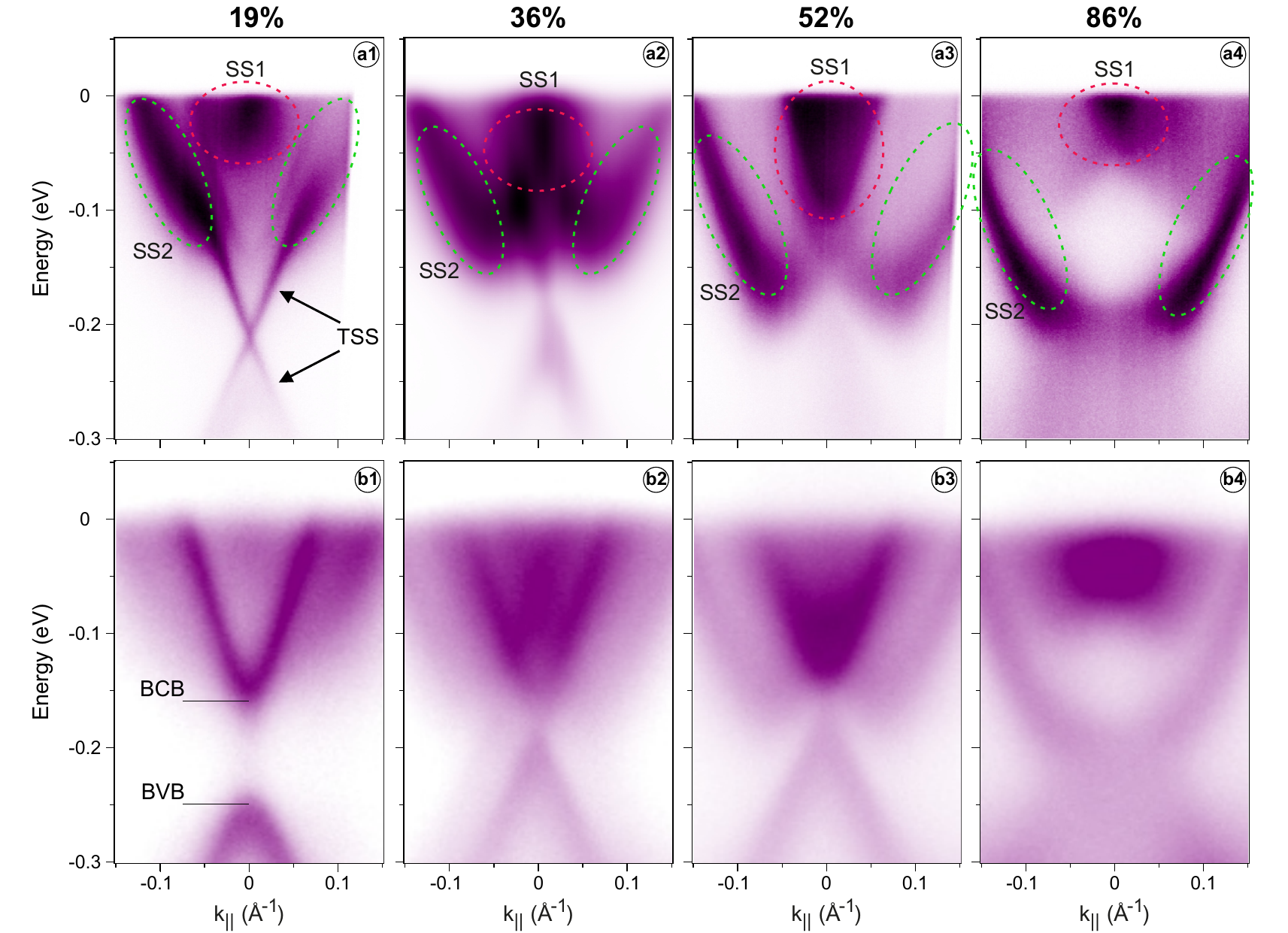}
    \caption{The results of ARPES for the \msbt{} system at various Sn concentrations: 19\% (\textbf{a1, b1}), 36\% (\textbf{a2, b2}), 52\% (\textbf{a3, b3}), and 86\% (\textbf{a4, b4}) are presented. The upper panel (\textbf{a1–a4}) shows data obtained using $\mu$-Laser ARPES ($h\nu$ = 6.3\ \text{eV}), while the lower panel (\textbf{b1–b4}) displays results acquired with He I$\alpha$ radiation ($h\nu$ = 21.2 \text{eV}).}
    \label{fig5}
\end{figure*}

Fig.~\ref{fig5} presents the results of ARPES measurements obtained using two different light sources. The upper panel shows the data acquired with the laser source, while the lower panel displays the results using He I$\alpha$ radiation. As demonstrated in previous studies~\cite{estyunina2023evolution, shikin2024phase}, utilizing $h\nu$ = 6.3~eV radiation allows for the visualization of TSS in \mbt{}, which are not visible with helium radiation. This distinction is clearly evident when comparing panels \textbf{a1} and \textbf{b1}, which show results for the system with 19\% tin. In panel \textbf{a1}, the TSS are well-defined, while in panel \textbf{b1}, they are conspicuously absent. At the same time, panel \textbf{b1} distinctly reveals the edges of the valence and conduction bands (VB and CB), which become visible in ARPES data due to n-doping caused by an excess of Te and Te/Bi antisite defects, characteristic of the sample synthesis process (for comparison, see ref.~\cite{Niu2012}).

Of particular interest for this system is the appearance of the SS1 and SS2 features, which we associate with the presence of RSS in the system. These features exhibit the highest intensity in Fig.~5(a1), where they hybridize with the TSS cone. In contrast, in Fig.~5(b1), the RSS appear more diffuse.

As the Sn concentration increases to 36\% (Fig.~\ref{fig5}), the panel (\textbf{a2}) shows that the TSS cone becomes less distinct, while the RSS features become more prominent, shifting downward in energy as predicted by theoretical calculations. On panel \textbf{b2}, the SS1 and SS2 are still less pronounced compared to those in the upper panel, highlighting the enhanced visibility of these features with $\mu$-Laser ARPES.

For the sample with 52\% Sn, which is calculated to be trivial, the disappearance of the TSS can be observed in Fig.~\ref{fig5}(a3). At the same time, SS1 and SS2 become most prominent at this concentration, shifting further downward in energy. The changes in Fig.\ref{fig5}(b3) compared to the previous panels of the lower row are particularly significant: at this concentration, the features of RSS become discernible. Additionally, the transition of the system to the trivial state is further corroborated by the closure of the energy gap, as observed in Fig.~\ref{fig5}(b3).

As the Sn concentration increases to 86\%, a state resembling the pure \sbt{} phase emerges, with the SS2 becoming most prominentand and the SS1 becoming barely distinguishable, while the TSS reappear in the electronic structure Fig.~\ref{fig5}(a4). A similar pattern was observed in theoretical calculations, attributed to the strong hybridization of the inner parts of RSS with other electronic states. The downward energy shift of the SS2 features continues. Fig.~\ref{fig5}(b4) at this concentration also yields intriguing results: it clearly reveals parts of RSS observed in Fig.~\ref{fig5}(a4). 
The dispersion pattern in panel Fig.\ref{fig5}(b4) is nearly identical to ARPES measurements for pristine \sbt{}\cite{Fragkos_2021, Li_2021, Eremeev2023}. In refs.\cite{Fragkos_2021, Li_2021}, the observed pattern of surface states was interpreted as TSS characteristic of \sbt{}. In ref.\cite{Eremeev2023}, it was noted that slightly above the TSS, RSS are present (referred to here as SS1). However, the results shown in panel Fig.~\ref{fig5}(a4) indicate that SS2 exhibit characteristics distinct from those of TSS, as they are observed only when using laser radiation. Thus, the experimental results confirm the main conclusions of the theoretical calculations, emphasizing the significant hybridization between the TSS and RSS.

\subsection{Theoretical insights into RSS}

\begin{figure*}[ht!]
    \centering
    \includegraphics[width=1.0\textwidth]{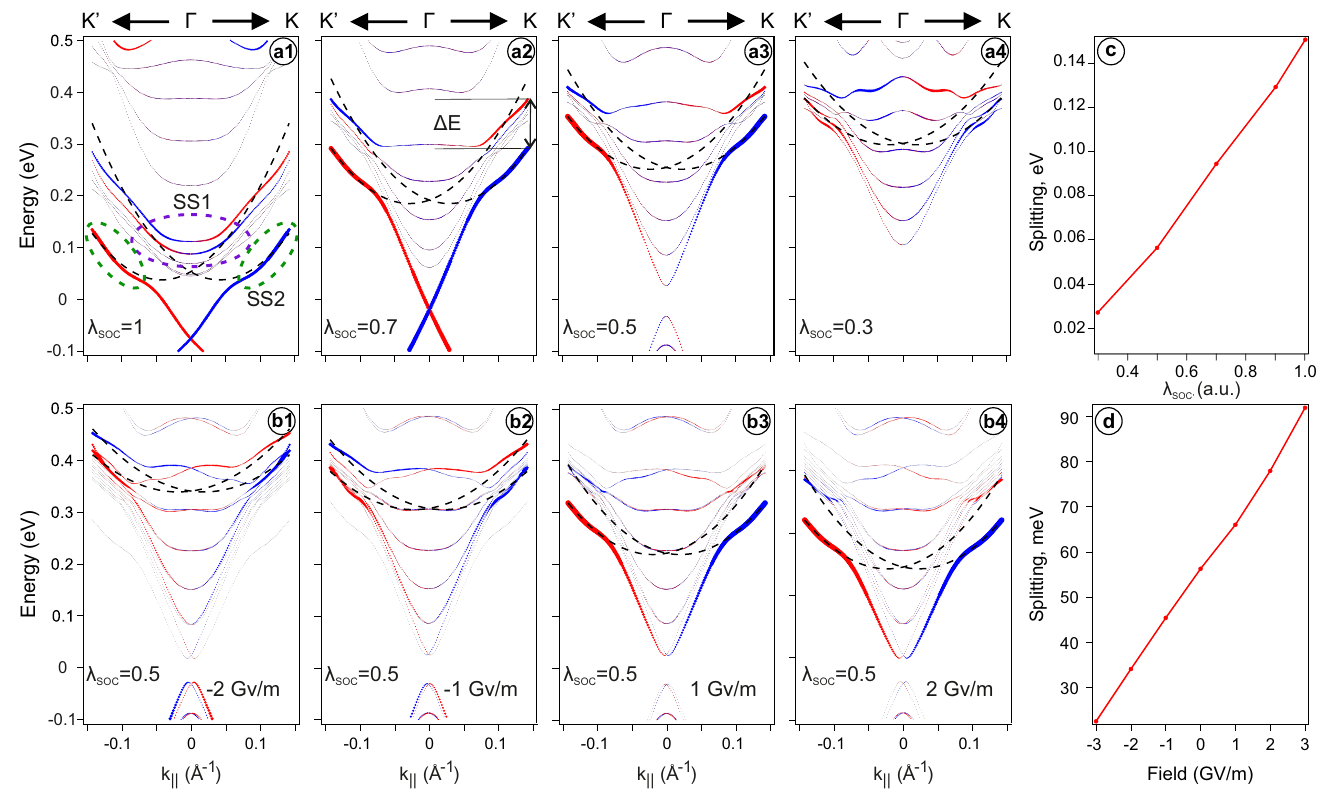}
    \caption{(\textbf{a1–a4}) Energy dispersion relations for the surface states localized in the first SL of \sbt{} at various values of spin-orbit coupling ($\lambda_{SOC}$). (\textbf{b1–b4}) Energy dispersions at a fixed $\lambda_{SOC}=0.5$ for different values of the external electric field. The plots on the right illustrate the dependence of the spin splitting value on $\lambda_{SOC}$ (\textbf{c}) and the external electric field (\textbf{d}).}
    \label{fig6}
\end{figure*}

Although the discussion so far has focused on RSS (based on the strong similarity between the observed states and typical Rashba-like features), the origin of these states requires further investigation. To uncover the mechanisms driving RSS, we examine their dependence on SOC strength, external electric fields, and surface charge in both pristine \mbt{} and \sbt{} crystals. Additionally, we analyze the localization and orbital composition of RSS under varying Sn concentrations, identifying trends that suggest a possible role for the orbital Rashba effect in enhancing spin-momentum locking. These findings aim to deepen our understanding of RSS and provide insights into their tunability for potential applications in spintronic devices.

\subsubsection{RSS in \texorpdfstring{\sbt{}}{sbt}}

In various studies on \sbt{}, alongside TSS, the presence of RSS in its electronic structure has already been noted~\cite{Eremeev2023}, which have been observed in other TIs with vdW-layered atomic structure~\cite{souma_topological_2012, bianchi_robust_2012, nomura_relationship_2014, benia_reactive_2011, zhu_rashba_2011, papagno_multiple_2016, pacile_deep_2018}. However, in all these works, RSS was understood as the feature we identified as SS1, and no investigation of the hybridization of RSS and TSS, manifested as the feature SS2, was conducted. It is necessary to fill this gap by proving the RSS nature of the SS2 feature.

To address it, Fig.~\ref{fig6} illustrates the dependence of the splitting magnitude of the studied states on two key factors: spin-orbit coupling strength and the magnitude of an external electric field to \sbt{}, which is chosen as the initial focus due to the relatively well-preserved RSS under hybridization effects.
These parameters were chosen to verify that the observed states exhibit characteristics consistent with Rashba-like physics, as spin-orbit coupling drives their formation, while the electric field should enable tuning of their energy splitting for further confirmation.

Fig.~\ref{fig6}(a1-a4) shows the impact of varying the spin-orbit coupling (SOC) strength. As noted earlier, the upper state in the pair of states labeled SS1 in Fig.~\ref{fig3}(a1) exhibits significantly higher localization in the first SL compared to the lower state, indicating their different origins. This conclusion is further supported by varying the SOC strength. It is clearly observed that when spin polarization of the 1SL states is considered, reducing the SOC strength to $\lambda_{SOC}=0.7$ causes the lower state to disappear from the overall spin texture. Upon further reduction of $\lambda_{SOC}$ to 0.5, the TSS also vanish from the spin texture. At the same time, the RSS remain visible at considered all SOC values.

From this, it can be concluded that the spin polarization of the lower state in the SS1 pair is associated with hybridization with the TSS, while the upper state belongs to the RSS. Furthermore, tracking the splitting of the RSS, identified based on these observations, reveals that the splitting magnitude decreases almost linearly with a reduction in the spin-orbit interaction strength (Fig.~\ref{fig6}(c)). Here, the splitting magnitude was measured as the energy difference between the two branches of the presumed RSS with opposite spin polarization.

\begin{figure*}[ht!]
    \centering
    \includegraphics[width=16cm]{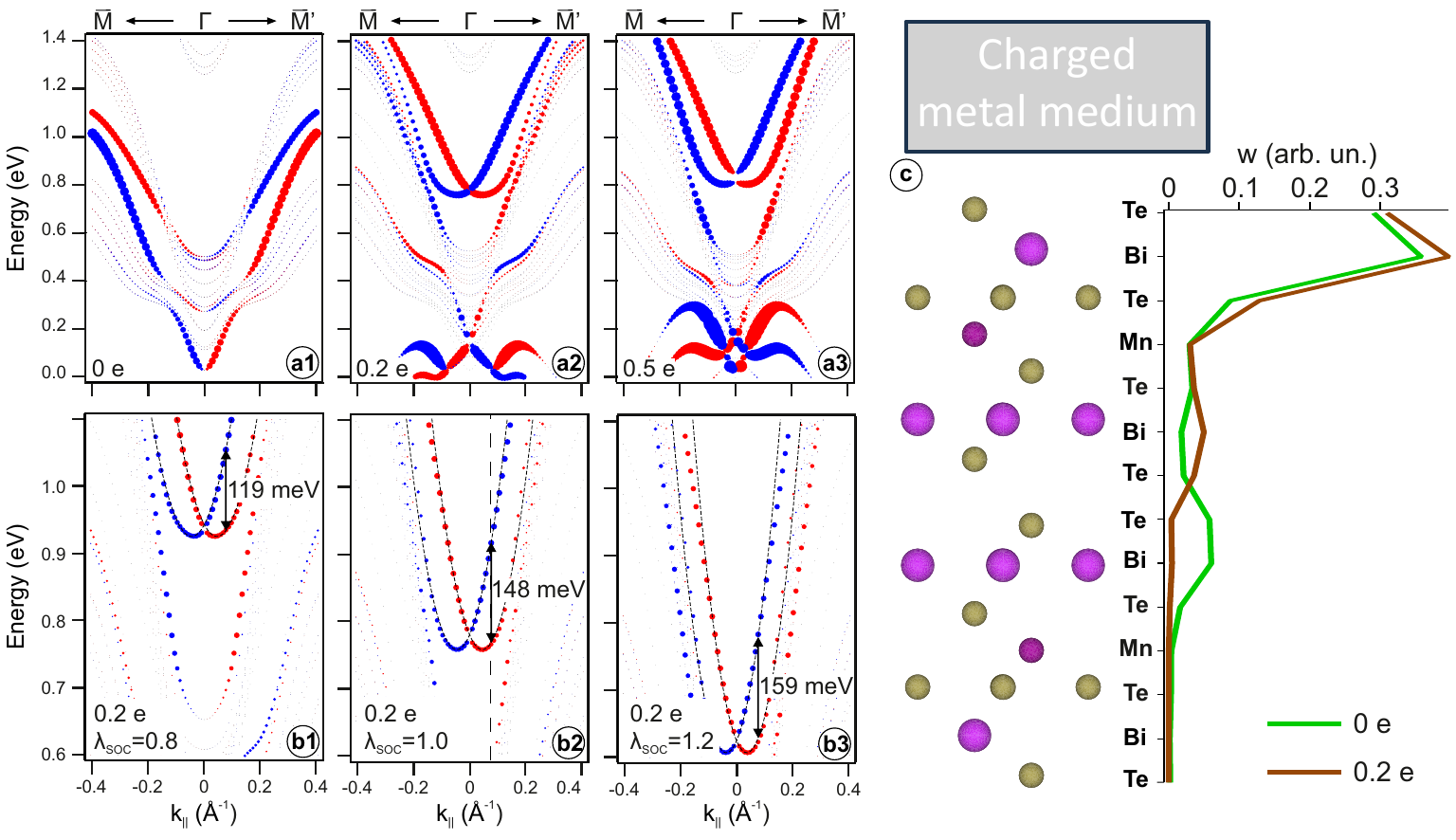}
    \caption{ (\textbf{a1–a3}) Energy dispersion relations for the surface states localized in the first SL of \mbt{} crystal with surface charge applied via a charged metallic environment. (\textbf{b1–b3}) A more detailed structure of the RSS under varying spin-orbit coupling strength at a fixed charge of 0.2 \textbf{e}. \textbf{(c)} Spatial distribution in the first two SLs of \mbt{} in the absence of an external charge (green curve) and with an applied charge of 0.2 \textbf{e} (brown curve) for the corresponding RSS (\textbf{a1, a2}), extracted from the overall electronic structure.}
    \label{fig7}
\end{figure*}

Thus, in order to further investigate the RSS, we can simplify the spin texture by removing the TSS through a reduction of $\lambda_{SOC}$. Fig.~\ref{fig6}(b1-b4) depicts the behavior of the system at $\lambda_{SOC}=0.5$ under an applied external electric field. The splitting magnitude exhibits the expected linear increase with growing field strength (Fig.~\ref{fig6}(d)), as expected for systems known to exhibit RSS. The observed linear relationship serves as direct evidence that the studied system demonstrates hallmark features of the Rashba-like physics, validating the accuracy of the chosen model. In addition to the states of the first SL highlighted in the figure, spin splitting caused by the transverse electric field also affects deeper states, resulting in the lifting of their spin degeneracy, as observed for the states not highlighted in Fig.~\ref{fig6}(b1-b4).

The analysis of \sbt{} confirms the presence of RSS within its electronic structure, distinguishing two key features: SS1, traditionally identified as RSS in prior studies, and SS2, a manifestation of RSS-TSS hybridization. By examining the effects of spin-orbit coupling strength and an external electric field, the hallmark behaviors of RSS, such as persistence under varying SOC and linear splitting dependence on electric field strength, were validated. These findings establish that the SS1 feature comprises two components with different origins: the upper state is predominantly Rashba-like, while the lower state arises from hybridization with TSS. The linearity of the splitting magnitude with respect to these parameters not only supports the Rashba nature of the states but also highlights the robustness of the theoretical model in capturing these phenomena. This understanding lays a critical foundation for exploring RSS tunability and its interplay with TSS in similar topological materials.

\subsubsection{RSS in \texorpdfstring{\mbt{}}{MBT}}

Next, we aim to confirm the presence of RSS in \mbt{}, even in the absence of Sn atoms. As noted earlier, the manifestation of RSS in \mbt{} has been reported in many experimental studies. However, it has not been consistently reproduced in theoretical calculations~\cite{otrokov2019prediction, hao_gapless_2019, swatek_gapless_2020}.

Fig.~\ref{fig7} presents the results of modeling the behavior of RSS in a pure \mbt{} crystal under an applied surface charge. Similar calculations of surface charge application to \mbt{} crystal were performed in \cite{shikin2021sample}, where it was demonstrated that the gap in TSS is influenced by the applied surface charge and can even be closed. However, that study did not investigate the impact of the applied charge on RSS. 

Fig.~\ref{fig7}(a1-a3) shows calculated band structures of \mbt{} where surface charge is gradually applied, increasing from 0~$e$ to 0.5~$e$. As the surface charge increases, the RSS experience upward energy shift and become more clearly visible. This suggests that RSS are indeed present in pure \mbt{}, however, their typical spin-momentum locked structure does not manifest itself clearly without a surface charge due to bulk hybridization. Nevertheless, the latter may be greatly suppressed by simple application of a sufficient surface charge which moves the RSS out of hybridization region, allowing them to be seen clearly.

In Fig.~\ref{fig7}(b1-b3), the $\lambda_\text{SOC}$ dependence of the spin splitting magnitude between different branches of RSS is examined after application of a 0.2~$e$ surface charge. The splitting magnitude increases almost linearly with the spin-orbit coupling constant ($\approx 29$~meV per each 0.2 increase in $\lambda_\text{SOC}$). This observation is similar to the behavior previously observed for \sbt{} without any additional surface charge. This consistency of linear splitting dependencies on $\lambda_\text{SOC}$ supports the suggestion that RSS are indeed a consequence of a strong spin-orbit interaction present in the studied systems. The reason behind the fact that various experimental evidence of RSS in \mbt{} were not supported by theoretical calculations may be a significant surface charge on the studied \mbt{} crystals.

Finally, Fig.~\ref{fig7}(c) shows that RSS become more surface-localized when an additional electric charge of appropriate sign is introduced into the system. This happens likely because the electrostatic potential near the surface is modified so that its gradient becomes shallower, allowing the RSS wave functions to occupy the resulting potential well near the surface more efficiently. This effect is also observed in similar systems with strong spin-orbit coupling~\cite{Ast_2007} and probably may be used to control the RSS energy position in and their spatial distribution in \mbt{} in a precise and practical manner. This offers valuable opportunities for their use in advanced spintronic devices.

\subsubsection{Localization and orbital contributions of RSS in \texorpdfstring{\msbt{}}{msbt}}

Spatial and orbital characteristics of RSS provide valuable insights into their formation mechanisms and dependence on material composition. In this section, we focus on Sn-enriched \msbt{} systems, as in these systems the RSS components, previously identified as SS2, do not overlap with the bulk CB states. This allows for a more precise analysis of the direct influence of changes in the atomic and orbital contributions on RSS localization, providing a clearer understanding of how these states evolve with increasing Sn concentration.

\begin{figure}
    \centering
    \includegraphics[width=\columnwidth]{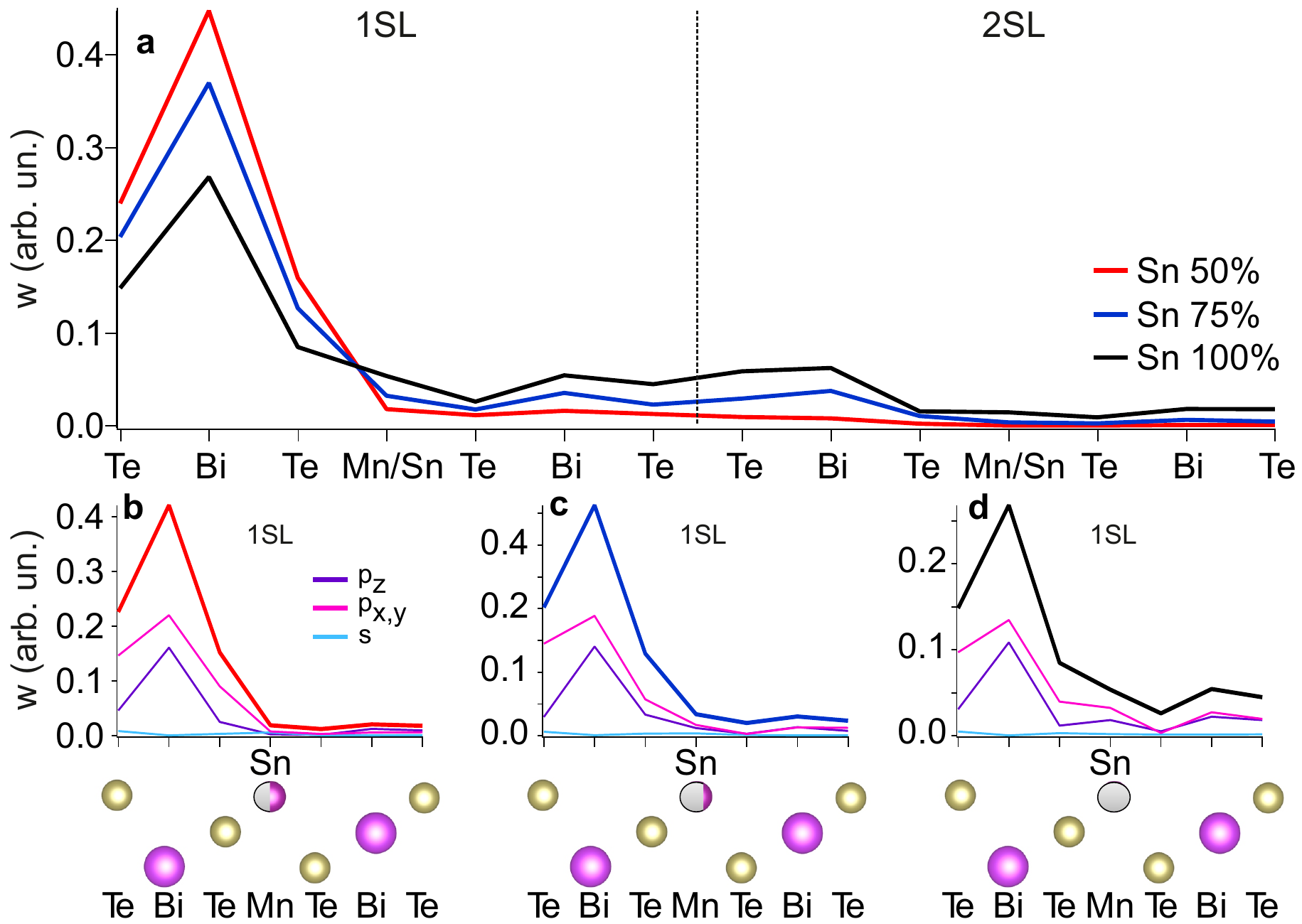} 
    \caption{Dependence of the localization of RSS in \msbt{} on Sn concentration (\textbf{a}), along with the decomposition of the total density of states into orbital components ($p_{x, y}$, $p_z$, and $s$) for different Sn concentrations: 50\% (\textbf{b}), 75\% (\textbf{c}), and 100\% (\textbf{d}). The bottom row illustrates the schematic arrangement of atoms in the first SL of the crystal.}
    \label{fig8}
\end{figure}

Fig.~\ref{fig8} shows RSS localizations through their averaged atomic and orbital compositions in SS2 regions of Fig.~\ref{fig2} for Sn concentrations (50\%, 75\%, and 100\%). It is evident that RSS are primarily localized in the first three Te--Bi--Te surface trilayer where Bi contribution from the second atomic layer dominates. RSS are composed mostly from $p$ orbitals; higher angular momentum orbital contributions are not shown since they vanish almost identically. Total atomic contributions and their $s$ and $p_z$ constituents are invariant with respect to rotations around the $z$ axis; hence the residue (symbolically designated as $p_{x, y}$) which corresponds to a planar contribution is also invariant and well-defined. One may see that this planar contribution is slightly greater than the $p_z$ contribution for all concentrations.

When Sn concentration increases past 75\%, it may be observed that RSS spread significantly deeper into the bulk up to the second Mn/Sn position whereas at lower concentrations RSS are mostly confined to the surface Te--Bi--Te trilayer (although the latter is harder to capture for concentrations below 50\% using atomic contributions since RSS are not so well separated from bulk states in this case). This behavior is likely connected to the fact that the valence shell of Mn consists of $s$ and $d$ electrons which have almost zero RSS contribution whereas that of Sn possess $p$ electrons as well which, according to Fig.~\ref{fig8}, take part in RSS formation more readily and thus facilitate movement of $p$ electrons past the first Sn position. Nevertheless, RSS are still predominantly localized near the surface even when Sn atoms fully substitute Mn atoms.

The slight dominance of $p_{x, y}$ contribution over that of $p_z$ may prove important in light of the so-called \enquote{orbital Rashba effect}, the notion used to distinguish between the conventional description of the Rashba effect based on surface potential gradients~\cite{bychkov1984properties,krasovskii2015spin,ast2007giant,premper2007spin} and another description explicitly involving atomic orbital hybridization mechanisms~\cite{bihlmayer2022rashba} to quantitatively capture anomalously strong Rashba effects which are not explained by the conventional scheme. Orbital Rashba effect requires that $s$ and $p$ orbitals hybridize in such a way that an effective in-plane, $\mathbf k$-antisymmetric angular momentum $\mathbf L(\mathbf k) = - \mathbf L(-\mathbf k)$ emerges which facilitates the distinctive spin-momentum locking feature of RSS~\cite{bihlmayer2022rashba,park2011orbital,go2017toward}. In this context, significantly pronounced character of RSS bands (where band dispersion aligns better with typical parabolic shape and its spin polarization becomes greater) at higher Sn concentrations correlates well with significant increase of Sn $p_{x, y}$ contributions into RSS which may suggest that the orbital Rashba effect indeed takes place in \msbt{} compounds. 

\section{Conclusion}

This work presents a comprehensive study of \msbt{} systems, emphasizing the tunability of their electronic structure through Sn doping and external parameters. High-quality polycrystalline and single-crystal \msbt{} samples were synthesized using the Bridgman method. X-ray diffraction confirmed their crystallization in the $\mathrm{Ge} \mathrm{As}_2 \mathrm{Te}_4$ structural type, with linear lattice parameter changes consistent with ’s law, highlighting precise compositional control and structural robustness.

DFT calculations reveal that the surface electronic structure of \msbt{} near the Fermi level is primarily shaped by the interaction between TSS and RSS. This interaction manifests in the formation of the SS1 and SS2 features, with SS1 retaining $Te-p$ orbital dominance and SS2 reflecting $Bi-p$ contributions across all Sn concentrations. Additionally, unlike SS1, SS2 exhibits strong localization in the first SL. The analysis highlights the intricate evolution of the electronic structure as the Sn concentration varies, emphasizing the interplay between RSS, TSS, and their hybridization effects.

At low Sn content, the system exhibits TSS with a gap, which vanish at 50\% Sn, signaling a topological phase transition to a trivial gapless state dominated by RSS. Beyond 75\%, the bulk band gap reopens, and TSS with a gap reemerge, significantly modified by their interaction with RSS. These transitions, induced by Sn doping, reflect a progression through intermediate phases, culminating in a nonmagnetic topological insulator state at 100\% Sn concentration. Spin texture analysis confirms the presence of RSS at all doping levels, strongly hybridized with TSS starting from low Sn content. At higher Sn concentrations, the hybridization of RSS with TSS becomes more pronounced due to the downward energy shifts of RSS, driven by doping-induced changes in the electronic structure.

Experimental ARPES measurements validate the theoretical predictions regarding the evolution of electronic states in \msbt{} systems with increasing Sn concentration. At low Sn concentrations (19\%), $\mu$-Laser ARPES reveals well-defined TSS, while He I$\alpha$ ARPES highlights the conduction and valence band edges influenced by n-type carriers. RSS features, hybridized with TSS, gain prominence as Sn concentration increases, shifting downward in energy and becoming dominant by 52\%, marking the transition to a trivial gapless state. At 86\%, the system resembles the pure \sbt{} phase, with reemerging TSS strongly hybridized with RSS. These observations align with theoretical predictions, confirming the critical role of Sn doping in shaping the electronic structure.

In \sbt{}, the presence of RSS is confirmed with two distinct features: SS1, traditionally identified as RSS, and SS2, arising from RSS-TSS hybridization. The robustness of RSS is validated through its persistence under varying spin-orbit coupling strength and its linear splitting dependence on external electric fields. These findings reveal that SS1 comprises two components: an upper state with a Rashba-like nature and a lower state influenced by TSS hybridization. This establishes a solid theoretical foundation for exploring RSS tunability and its interplay with TSS in other topological materials.

In \mbt{}, RSS are observed even in the absence of Sn, with their upward energy shift and enhanced prominence under increasing surface charge providing further evidence of their Rashba-like form. Surface charge application demonstrates precise control over RSS localization and energy position, pulling their wave functions closer to the surface due to the altered electrostatic potential. Additionally, the splitting magnitude of these states remains linearly proportional to spin-orbit coupling strength, further corroborating their Rashba-like characteristics.

In Sn-enriched \msbt{} systems, Sn doping profoundly influences the evolution and localization of RSS. These states are primarily confined to the first three Te--Bi--Te trilayers, with $Bi-p$ orbital dominance across all Sn concentrations. Beyond 75\% Sn, RSS extend deeper into the bulk, driven by the participation of $Sn-p$ orbitals, which contrasts with the negligible contribution of Mn’s $s$ and $d$ orbitals. The slight dominance of the $p_{x, y}$ orbital component over $p_z$ across all concentrations suggests the presence of the orbital Rashba effect, which may account for the enhanced spin-momentum locking and pronounced RSS characteristics observed at higher Sn concentrations.

In summary, this study provides a comprehensive framework for understanding and controlling the electronic properties of \msbt{} systems through Sn doping and external parameters. The combined experimental and theoretical approach reveals a tunable interplay between TSS and RSS, driving a diverse range of electronic behaviors. These findings not only deepen our understanding of the fundamental mechanisms shaping these materials but also highlight their potential for advanced technological applications. By elucidating the evolution of electronic states under Sn doping and external influences, this work establishes a foundation for designing materials with finely tunable electronic characteristics, offering practical pathways for their integration into future quantum and spin-based technologies.

\begin{acknowledgments}
The authors acknowledge support by the Saint Petersburg State University (Grant No. 116812735), Russian Science Foundation (Grant No. 23-12-00016 in the part of  in the part of theoretical calculations and analysis of the electronic and spin structure and experimental ARPES measurements and Grant No. 22-72-10074 in the part of experimental and theoretical analysis of XRD and ESM calculations). A.S.F. and A.I.S. acknowledge support of Russian Science Foundation (Grant No. 23-72-00020) for their work on synthesis of polycrystalline \msbt{} samples.
 This work was supported by “Centre for X-ray Diffraction Studies” and “Centre for Nanotechnology”  of the Research Park of St. Petersburg University.
Monocrystalline \msbt{} samples were
grown using the Bridgeman method under the state assignment of Sobolev Institute of Geology and Mineralogy SB RAS No. 122041400031-2.
HiSOR ARPES measurements were performed under Proposals No. 23AG008 and 23AU003. We are
grateful to the N-BARD, Hiroshima University for liquid He supplies.  The
calculations were partially performed using the equipment of the Joint
Supercomputer Center of the Russian Academy of Sciences
(https://rscgroup.ru/en/project/jscc).
\end{acknowledgments}

\end{document}